\documentclass{article}
\usepackage{spconf,amsmath,graphicx}
\usepackage{enumitem}
\usepackage{url}
\usepackage{xcolor}

\title{Optimizing voice conversion network with cycle consistency loss of speaker identity}
\name{Hongqiang Du\textsuperscript{1,2}, Xiaohai Tian\textsuperscript{2}, Lei Xie\textsuperscript{1*}\thanks{*Lei Xie is the corresponding author, lxie@nwpu.edu.cn.}, Haizhou Li\textsuperscript{2}}
\address{
	\textsuperscript{1}Audio, Speech and Langauge Processing Group (ASLP@NPU), School of Computer Science,\\ Northwestern Polytechnical University, Xi'an, China \\
	\textsuperscript{2}Department of Electrical and Computer Engineering, National University of Singapore, Singapore \\
	hongqiang.du@u.nus.edu, eletia@nus.edu.sg, lxie@nwpu.edu.cn, haizhou.li@nus.edu.sg
}
\begin{document}
%
\maketitle
\begin{abstract}
	
We propose a novel training scheme to optimize voice conversion network with a speaker identity loss function. The training scheme not only minimizes frame-level spectral loss, but also speaker identity loss. We introduce a cycle consistency loss that constrains the converted speech to maintain the same speaker identity as reference speech at utterance level. While the proposed training scheme is applicable to any voice conversion networks, we formulate the study under the average model voice conversion framework in this paper. Experiments conducted on CMU-ARCTIC and CSTR-VCTK corpus confirm that the proposed method outperforms baseline methods in terms of speaker similarity.

\end{abstract}
\section{Introduction}
\label{sec:intro}

Voice conversion (VC)~\cite{mohammadi2017overview} aims to modify a speech signal uttered by a source speaker to sound as if it is uttered by a target speaker while retaining the linguistic information. This technique has various applications, such as emotion conversion, voice morphing, personalized text-to-speech synthesis, movie dubbing as well as other entertainment applications.

A voice conversion pipeline generally consists of feature extraction, feature conversion, and speech generation. In this work, we focus on feature conversion. Many studies have been devoted to the conversion of spectral features between a specific source-target speaker pair, for example,  Gaussian mixture model (GMM)~\cite{toda2007voice,kain1998spectral,stylianou1998continuous,benisty2011voice}, frequency warping~\cite{erro2009voice,godoy2011voice,tian2015sparse,tian2014correlation}, exemplar based methods~\cite{takashima2012exemplar,wu2014exemplar,tian2017exemplar}, deep neural network (DNN)~\cite{desai2009voice,chen2014voice,mohammadi2014voice},  and long short-term memory (LSTM)~\cite{sun2015voice}. 

To benefit from publicly available speech data and to reduce the amount of required target data, average model based approaches are proposed. Instead of training a conversion model for target speaker from scratch, we first train a general model with a multi-speaker database, and then adapt the general model towards the target with a small amount of target data~\cite{tian2018average,liu2018wavenet,du2020effective}, that is referred to as the average model approach.  
Alternatively, in some other studies, a speaker vector, e.g. one-hot vector, i-vector, or speaker embedding, is utilized as an auxiliary input to control the speaker identity. As one-hot speaker vector only works for close-set speakers, for example, in variational auto-encoder (VAE)~\cite{hsu2016voice,van2017neural}, i-vector is a better speaker representation for unseen speakers~\cite{liu2018voice}. There are also other studies on speaker embedding techniques~\cite{zhou2019many,lu2019one,chou2018multi,chou2019one,zhang2019non} for voice conversion. 

Despite the progress, speaker similarity to target speaker of the above techniques remains to be improved~\cite{lorenzo2018voice}. One of the reasons is that such methods attempt to minimize the difference between the converted and target features at acoustic feature space, which is not directly related to the speaker identity. To further improve the speaker similarity between generated and target speech, recent studies propose a perceptual loss as a feedback constraint for speech synthesis. In~\cite{cai2020speaker}, a feedback constraint on the speaker embedding space is used for speech synthesis. The work~\cite{nakamura2019v2s} proposes a verification-to-synthesis framework, where VC is trained with an automatic speaker verification (ASV) network.

In this paper, we introduce a cycle consistency loss on speaker embedding space to enhance the speaker identity conversion for the average model VC approach. In the proposed approach, the speaker independent phonetic posteriorgrams (PPG)~\cite{sun2016phonetic} are used to represent the content information, while a speaker embedding extracted from a pre-trained speaker embedding extractor is used to control the generated speaker identity. To ensure that the generated speech 
preserves the target speaker identity, the cycle consistency loss encourages the speaker embedding of the converted speech is the same as the input speaker embedding.

The rest of this paper is organized as follows. Section~\ref{sec:AMA}
briefly describes the average modeling approach for voice conversion. The details of our proposed method are discussed in Section~\ref{sec:AMA_SV}. The experimental setup and results are presented in Sections~\ref{sec:setups} and ~\ref{sec:evaluations} respectively. Finally, Section~\ref{sec:conclusion} concludes the study.
\vspace{-0.1cm}

\section{Average modeling approach for voice conversion}
\label{sec:AMA}

\begin{figure}[!ht]
	\centering
	\centerline{\includegraphics[width=\linewidth]{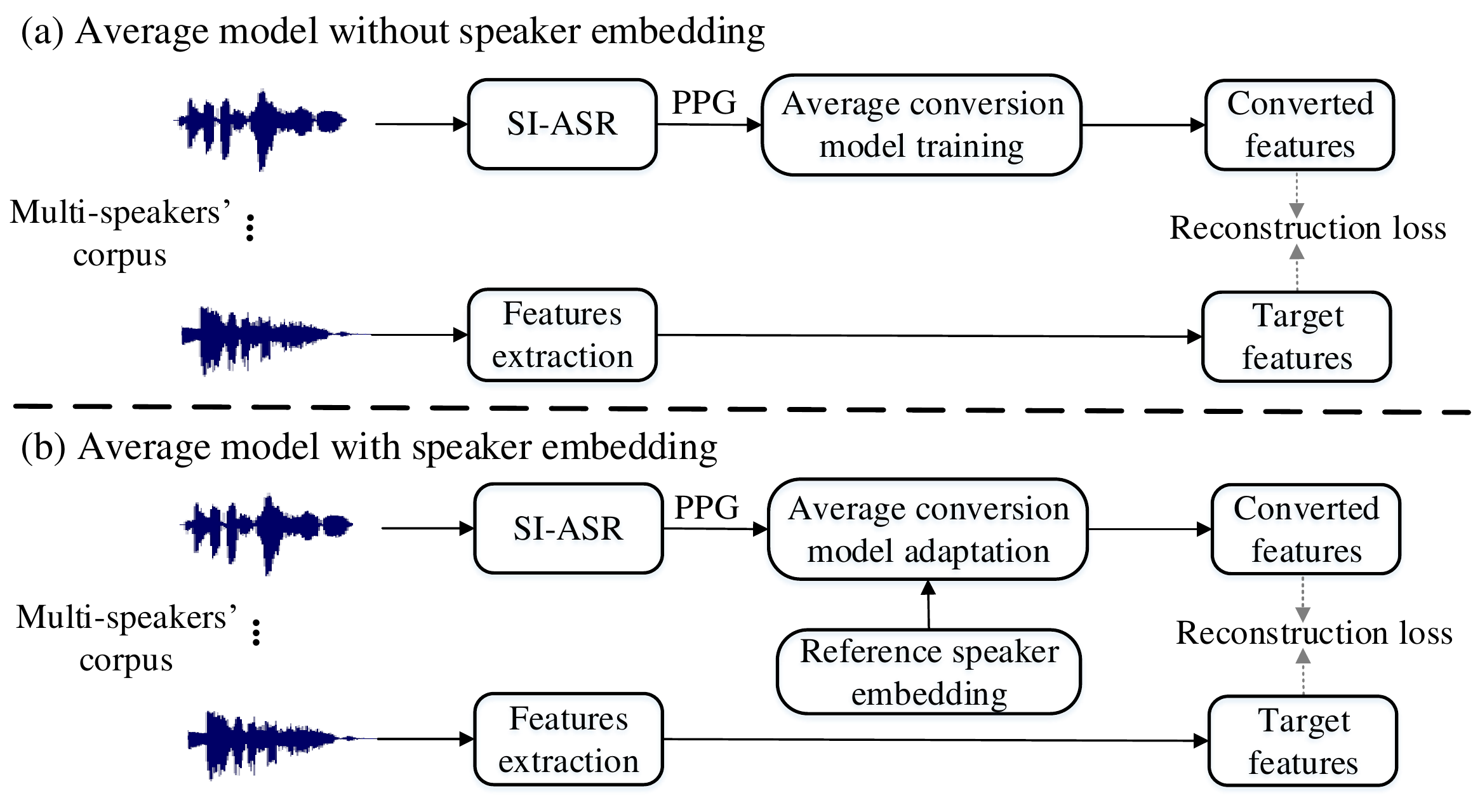}}

	
	\caption{Diagram of the average modeling approach for voice conversion. Fig (a) shows the training process of average model without speaker embedding as input. The average model is optimized with reconstruction loss in feature space. Fig (b) shows the training process of average model conditioned on target speaker embedding.}
	\label{fig:AMA}

\end{figure}

A popular average modeling approach (AMA) makes use of speaker independent PPG features~\cite{tian2018average,liu2018wavenet,du2019wavenet} as the feature representation, that allows us to use multi-speaker, publicly available data for voice conversion modeling.  It takes PPG features as input and generates mel-cepstral coefficients(MCCs). Fig.~\ref{fig:AMA} (a) presents the training process of average model without speaker embedding as input. In the adaptation phase, the conversion model is fine-tuned with a small number of target data.  Fig.~\ref{fig:AMA} (b) presents the training process of average conversion model with speaker embedding as input. Unlike Fig.~\ref{fig:AMA} (a), Fig.~\ref{fig:AMA} (b) uses a speaker embedding to control the speaker identity. During adaptation, the speaker embedding is extracted from target speech, and the model is also fine-tuned with target data.

For both AMA with and without speaker embedding, the spectral distortion loss is used for network optimization. We note that spectral distortion loss aims to minimize the frame-level loss, that does not directly reflect the utterance level speaking style and speaker characteristics. According to the results in VCC 2018~\cite{lorenzo2018voice}, there still exists a gap in terms of speaker similarity in speaker identity space. 

\section{Speaker identity conversion with cycle consistency loss}
\label{sec:AMA_SV}

\subsection{Cycle consistency loss}
Cycle consistency generally assumes reversibility between source domain and target domain, e.g. domain A can be translated to domain B and vice-versa to maintain the same content information. Cycle consistency loss is first studied in~\cite{zhu2017unpaired} for image-to-image translation, which is effective in maintaining the same image content during translation. The same idea is used in voice conversion~\cite{kaneko2017parallel}. Recently, a style cycle consistent loss~\cite{whitehill2019multi} is proposed. 
Similarly, a linguistic cycle consistent loss~\cite{luong2020nautilus,qian2019autovc} is used to maintain the same linguistic when translating only from source to target by using reference information. 

\subsection{Cycle consistency loss of speaker identity}

We propose a novel loss function, cycle consistency loss, to improve the speaker identity conversion for average model voice conversion. PPG based average modeling approach is used in our system with a speaker embedding as an auxiliary input to control the speaker identity of output speech. Unlike previous studies~\cite{luong2020nautilus,qian2019autovc}, the cycle consistency loss ensures that the converted speech is close to the target in speaker embedding space, that characterizes speakers at utterance level. It is similar to perceptual loss in image reconstruction\cite{Johnson2016Perceptual}.

Speaker embedding compresses an arbitrary length of utterance to a fixed dimensional vector in speaker identity space. It is expected to capture high level speaker characteristics that are independent of the phonetic information~\cite{wan2018generalized}. 

\begin{figure*}[!ht]
	\centering
	\centerline{\includegraphics[width=\linewidth]{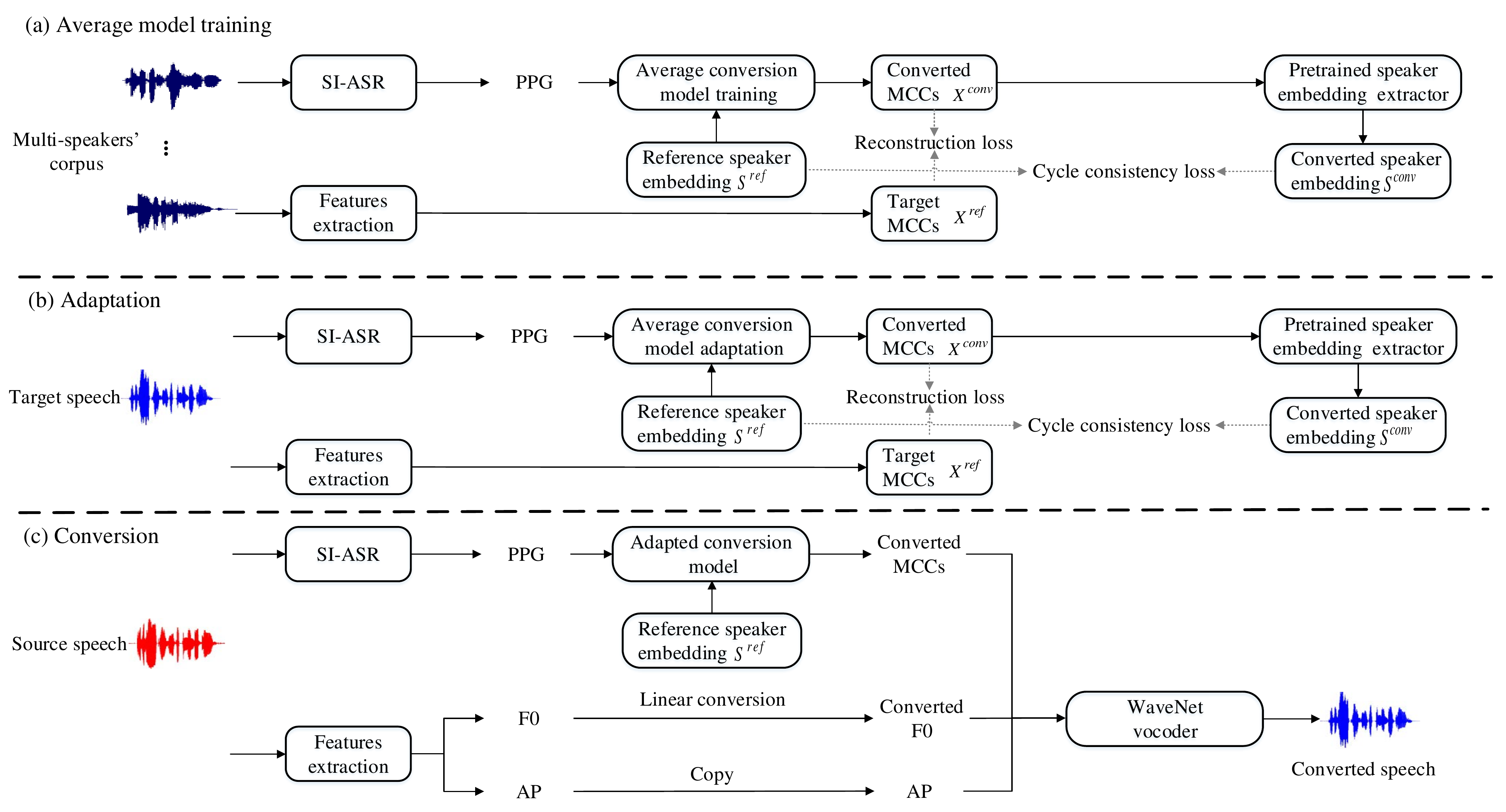}}
	\vspace{-0.3cm}
	
	\caption{Diagram of the average modeling approach with cycle consistency loss for voice conversion. Fig (a) shows the training process of average conversion model. It is optimized with reconstruction loss in feature space and cycle consistency loss in speaker embedding space. Fig (b) presents the adaptation process of average conversion model with two loss functions. Fig (c) shows the run-time conversion process.}
	\label{fig:AMA_SV}
\end{figure*}

The training procedure is illustrated in Fig.~\ref{fig:AMA_SV} (a) and (b). An average model is first trained with a multi-speaker corpus. Then a small amount of target data is used to adapt the average model towards the target speaker. In both training and adaptation, reconstruction loss and cycle consistency loss are used to optimize voice conversion model. We now explain the two loss functions.

\textbf{Reconstruction Loss}. Let ${X^{conv}}$ and ${X^{ref}}$ denote a converted MCCs and its target MCCs respectively. At frame-level, we use ${X^{ref}}$ as the learning target of ${X^{conv}}$. The reconstruction loss of $N$ dimensional MCCs is expressed as:

\begin{equation}\label{eq:rec}
	{L_{rec}} = {\sum^N_{n=1} {\left( {{X^{conv}_n} - {X^{ref}_n}} \right)} ^2}.
\end{equation}

\textbf{Cycle Consistency Loss}. The reconstruction loss is intended for optimizing output of individual frames. It is not designed to optimize speaking style and speaker identity at utterance level. We follow the idea of perceptual loss in image transformation~\cite{Johnson2016Perceptual} and introduce a cycle consistency loss that compares high level speaker identity features.  Let ${{S^{ref}}}$ be the reference speaker embedding and ${{{{S}}^{conv}}}$ the converted speaker embedding extracted from a pre-trained speaker embedding extractor, that is referred to as the loss network. The pre-trained loss network allows us to compare ${{S^{ref}}}$ with ${{{{S}}^{conv}}}$ at utterance level. The speaker embedding cycle consistency loss is calculated as follows:
\vspace{-0.05cm}
\begin{equation}\label{eq:cc}
{L_{cc}} = \sqrt {\sum\limits_{m = 1}^M {{{(S_m^{conv} - S_m^{ref})}^2}} }.
\end{equation}

where $M$ is is the dimension of speaker embedding. Note that the parameters in speaker embedding extractor are not updated during training process. ${L_{rec}}$ and ${L_{cc}}$ are combined as a joint loss function to optimize the voice conversion model during training. The overall loss function is thus given as follows:

\begin{equation}\label{eq:all}
{L_{all}} = {L_{rec}} + \alpha {L_{cc}},
\end{equation}
where $\alpha$ is a hyper-parameter, specifying the weight of ${L_{cc}}$ to balance the two losses.

At run-time conversion, we first feed  PPG of source speech as  input  to obtain converted MCCs. Then a linear transformation~\cite{toda2007voice} is performed to obtain the converted $f0$. These converted features combined with source aperiodicity (AP) are used as the input of WaveNet vocoder for speech generation~\cite{sisman2018voice}.

\section{Experiments setups}
\label{sec:setups}

\subsection{Database and feature extraction}

CSTR-VCTK~\cite{veaux2017cstr} database, containing 44 hours of speech from 109 speakers, is used to train average conversion model and speaker embedding extractor. CMU-ARCTIC~\cite{kominek2004cmu} database is used to perform voice conversion. We select four speakers, including two female speakers (clb and slt) and two male speakers (bdl and rms). For each target speaker, we randomly select 50 sentences for average conversion model adaptation, while another 20 non-overlap utterances are used for evaluation. All audio files are downsampled to 16 kHz.

WORLD vocoder~\cite{morise2016world} is employed to extract 1 dimensional $f0$, 1 dimensional aperiodicity coefficient, and 513 dimensional spectrum with 5ms frame shift and 25 ms window length. Then we calculate 40 dimensional mel-cepstral coefficients (MCCs) from spectral by speech signal processing toolkit (SPTK). 42 dimensional phonetic posteriorgram (PPG) features are extracted by the SI-ASR system trained on the Wall Street Journal corpus (WSJ)~\cite{paul1992design}.

\subsection{Systems and setup}

\begin{itemize}[leftmargin=*]
	\itemsep0em
	\item \textbf{AMA-R}: This is the average modeling approach (AMA)~\cite{tian2018average} for voice conversion, which is optimized only with reconstruction loss. 
	\item \textbf{AMA-SE-R}: This has the same setting as AMA except that the speaker embedding is used as an auxiliary input of conversion model.
	
	\item \textbf{AMA-RC}: This has the same setting as AMA except that the average conversion model is optimized with reconstruction loss and cycle consistency loss.
	
	\item \textbf{AMA-SE-RC}: This has the same setting as AMA-RC except that we extract speaker embedding and feed it into voice conversion model as an auxiliary input. 

\end{itemize}

The AMA consists of one feed-forward layer and four long short-term memory layers. Each hidden layer consists of 256 units. The speaker embedding extractor is a residual network (ResNet) based model in our work. It contains 5 residual blocks, followed by a $1 \times 1$ convolution layer and a mean pooling layer. Each residual block consists of two $1 \times 1$ convolution layers and a 1-D max-pooling layer. A ReLU activation is added after each $1 \times 1$ convolution layer. A residual connection is used to add the input to the output of the second $1 \times 1$ convolution layer. The parameter $\alpha$ is set to 0.2.

WaveNet vocoder is used for speech signal reconstruction, which consists of 3 residual blocks and each block contains 10 dilated convolution layers. The filter size of causal dilated convolution is 2. The hidden units of residual connection and skip connection are 256 respectively. We train the WaveNet vocoder for 600,000 steps using the Adam optimization method with a constant learning rate of 0.0001. The mini-batch sample size is 14,000. The speech is encoded by 8 bits $\mu$-law.

\section{Evaluations}
\label{sec:evaluations}

\subsection{Objective evaluation}

Mel-cepstral distortion (MCD) is employed to measure how close the converted is to the target speech. MCD is the Euclidean
distance between the MCCs of the converted speech and the
target speech. Given a speech frame, the MCD is calculated as follows: 
\begin{equation}\label{eq:mcd}
\vspace{1mm}
\text{MCD[dB]} = \frac{{10}}{{\ln 10}}\sqrt {2\sum\limits_{n=1}^N {{{\left( {X_n^{conv} - X_n^{ref}} \right)}^2}} },
\end{equation}
where $X_n^{conv}$ and $X_n^{ref}$ are the $n^{th}$ coefficient of the converted and target MCCs, $N$ is the dimension of MCCs. The lower MCD indicates the smaller distortion.

We also employ Eq.~\ref{eq:cc} to compute the cycle consistency distortion (CCD) between the speaker embeddings of converted speech and target speech to measure the speaker similarity objectively.




Table~\ref{table:rmse} shows the average MCD and CCD results for different systems. Firstly, we examine the effect of 
cycle consistency loss for voice conversion model without speaker embedding. It is observed that AMA-RC outperforms AMA-R in both MCD and CCD respectively. Then, we validate the effect of cycle consistency loss when voice conversion model uses speaker embedding as an auxiliary input. We observe that AMA-SE-RC outperforms AMA-SE-R in terms of CCD, while performing similar to AMA-SE-R in terms of MCD. Finally, we compare the performance of voice conversion model with or without speaker embedding when optimized with cycle consistency loss. As speaker embedding controls the speaker identity and cycle consistency loss ensures the speaker information consistency between converted speech and target speech, AMA-SE-RC outperforms AMA-RC and achieves the lowest CCD of 1.23.


\begin{table}[]
	\renewcommand\arraystretch{1.4}
	\small
	\centering
	\caption{Comparison of average MCD (dB) and average CCD for different systems.}
	\setlength{\tabcolsep}{4.4mm}{
		\begin{tabular}{|c|c|c|}
			\hline
			Systems  & Average MCD & \begin{tabular}[c]{@{}c@{}}Average CCD
			\end{tabular} \\ \hline
			AMA-R      & 6.52        & 1.68                                                               \\ \hline
			AMA-RC   & 6.41        & 1.36                                                                 \\ \hline
			AMA-SE-R      & 6.20        & 1.52                                                               \\ \hline			
			AMA-SE-RC & 6.24        & 1.23                                                                 \\ \hline
		\end{tabular}}
	\label{table:rmse}
\end{table}

\subsection{Subjective evaluation}
For subjective evaluation, we first conduct AB and XAB preference tests to assess speech quality and speaker similarity. Then the mean opinion score (MOS) is utilized to evaluate our models. Each listener is asked to give an opinion score on a five-point scale (5: excellent, 4: good, 3: fair, 2: poor, 1: bad).
For each system, 20 samples are randomly selected from the 80 converted samples for listening tests. 20 English listeners participated in all listening tests. Different listeners may listen to different samples. Listening tests cover all the 80 evaluation samples. 
 
\begin{figure}[!ht]
	\centering
	\centerline{\includegraphics[width=1.0\linewidth]{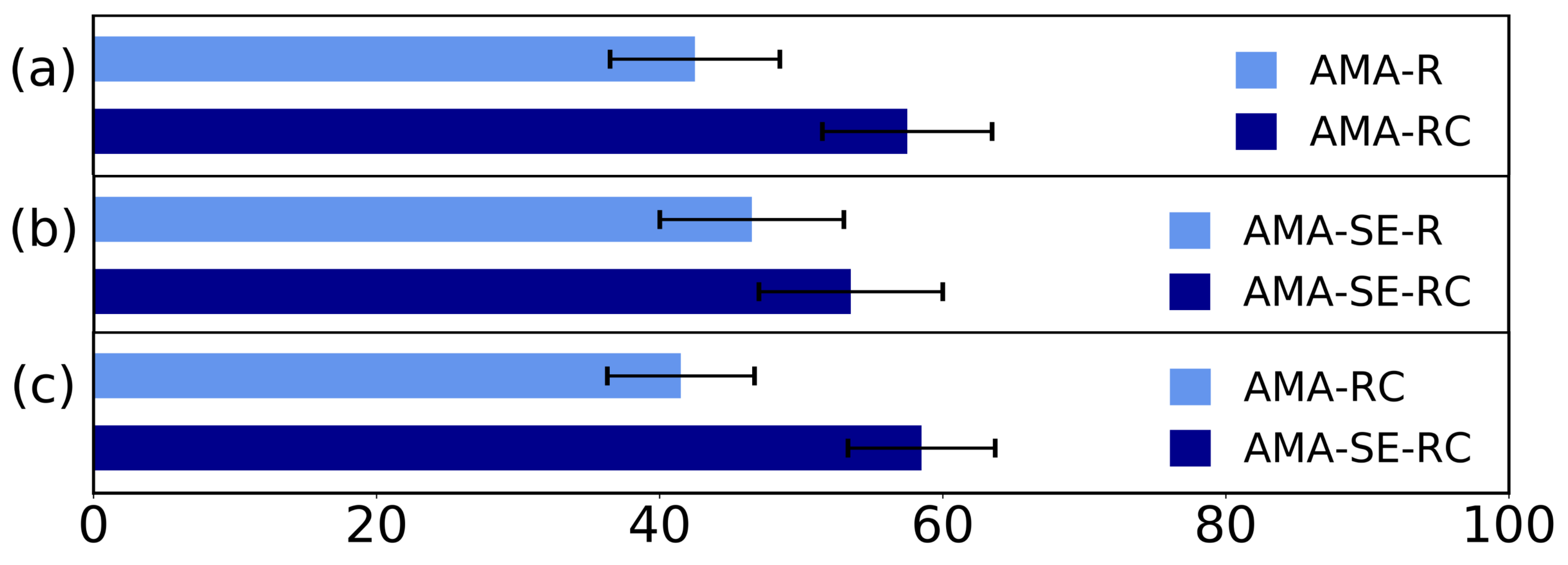}}
	
	\caption{Results of speech quality preference tests with 95\% confidence intervals for (a) AMA-R vs. AMA-RC, (b) AMA-SE-R vs. AMA-SE-RC, (c) AMA-RC vs. AMA-SE-RC.}
	\label{fig:ab_AMA_SV}
\end{figure}

\begin{figure}[!ht]
	\centering
	\centerline{\includegraphics[width=\linewidth]{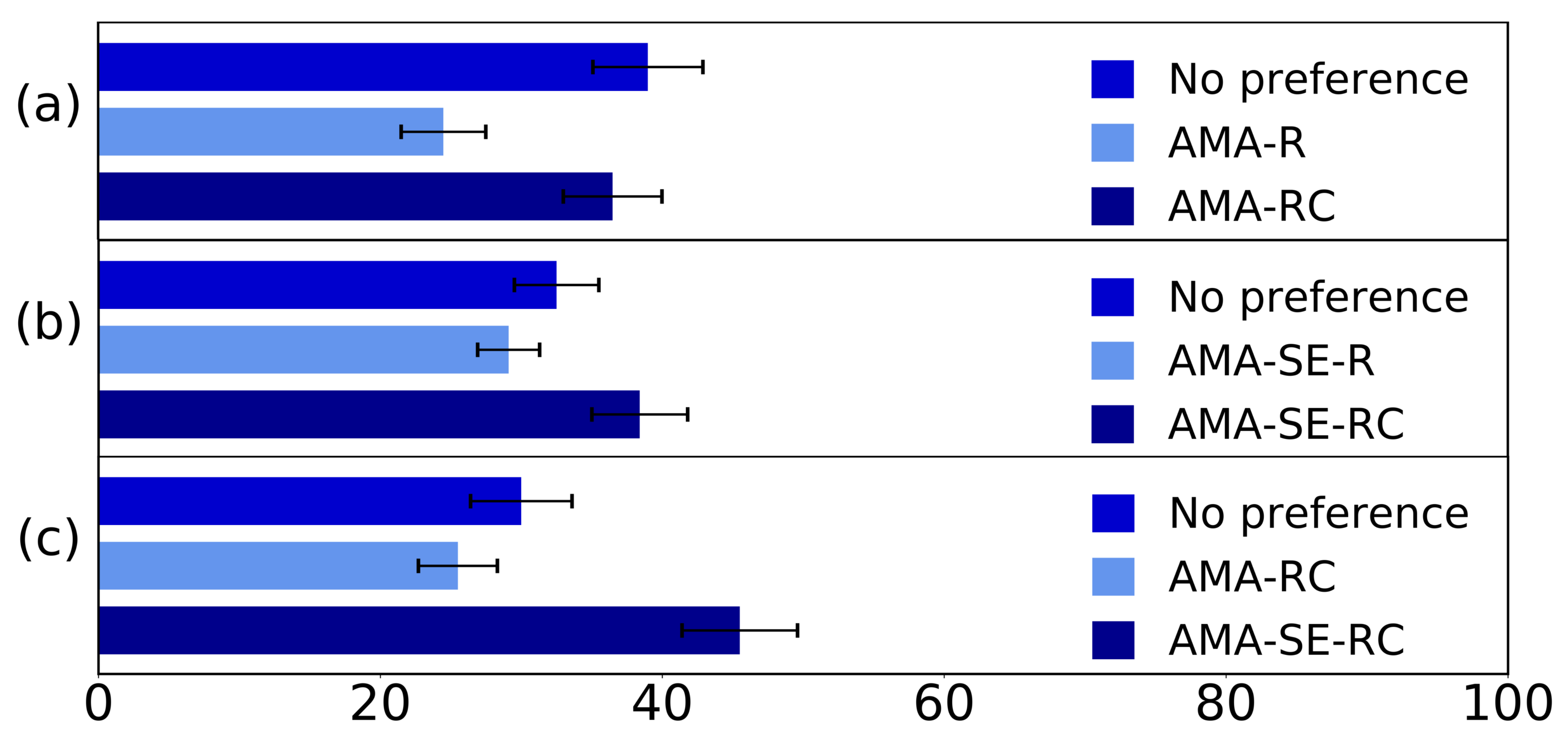}}
	
	\caption{Results of speaker similarity preference tests with 95\% confidence intervals for (a) AMA-R vs. AMA-RC, (b) AMA-SE-R vs. AMA-SE-RC, (c) AMA-RC vs. AMA-SE-RC.}
	\label{fig:xab_AMA_SV}
\end{figure}

The subjective results of AB tests are presented in Fig.~\ref{fig:ab_AMA_SV}. We first examine the effect of cycle consistency loss. As shown in Fig.~\ref{fig:ab_AMA_SV} (a), AMA-RC outperforms AMA-R in terms of speech quality. We further examine the effect of speaker embedding as input, we observe that AMA-SE-RC achieves results that are similar to AMA-SE-R as shown in Fig.~\ref{fig:ab_AMA_SV} (b). Fig.~\ref{fig:ab_AMA_SV} (c) shows that AMA-SE-RC outperforms AMA-RC, this validates the idea of speaker identity cycle consistency loss. It achieves the best results working together with speaker embedding input.

\begin{figure}[!ht]
	\centering
	\centerline{\includegraphics[width=0.95\linewidth]{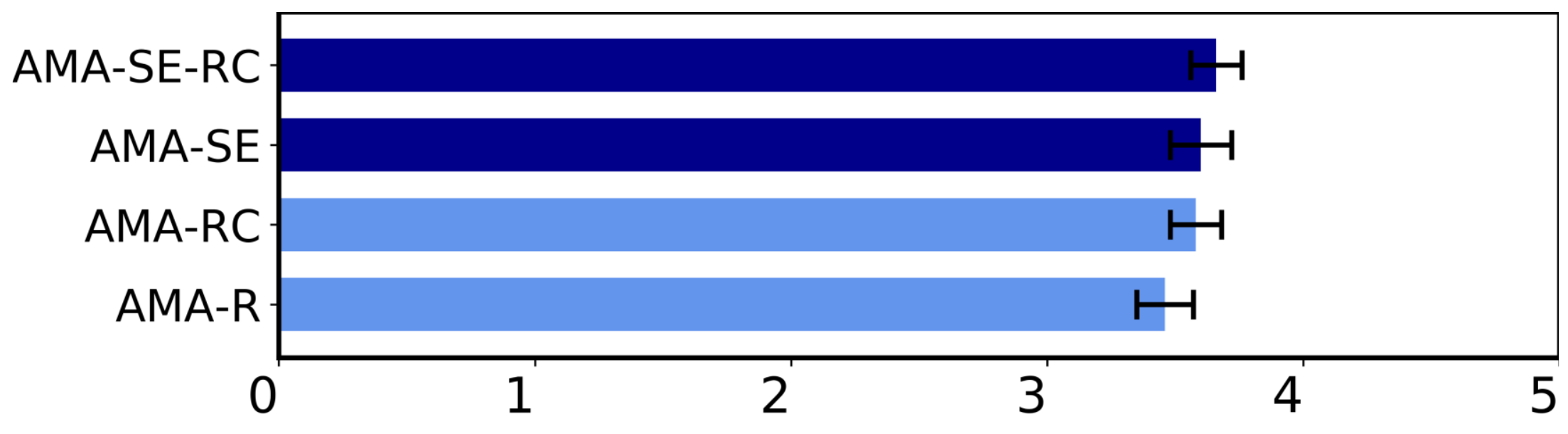}}
 	\vspace{-0.3cm}
	
	\caption{Comparison of mean opinion scores for different systems.}
	\label{fig:mos}
\end{figure}


Fig.~\ref{fig:xab_AMA_SV} shows the similarity preference results of XAB tests. As shown in Fig.~\ref{fig:xab_AMA_SV} (a), we observe that AMA-RC outperforms AMA-R in terms of speaker similarity due to the cycle consistency loss. Fig.~\ref{fig:xab_AMA_SV} (b) suggests that AMA-SE-RC also outperforms AMA-SE-R.  Fig.~\ref{fig:xab_AMA_SV} (c) shows that AMA-SE-RC performs better than AMA-RC, this indicates that speaker similarity can be further improved when speaker embedding is used as an auxiliary input for the conversion model optimized with cycle consistency loss.

Fig.~\ref{fig:mos} shows the mean opinion scores for different systems. Benefiting from the cycle consistency loss and speaker embedding as an auxiliary input for voice conversion model, AMA-SE-RC achieves the highest MOS score.

Therefore, we conclude that the speaker similarity of voice conversion can be improved by cycle consistency loss. The synthesized samples with different
systems can be found on the website~\footnote{\url{https://dhqadg.github.io/CCL/}}.

\section{Conclusion}
\label{sec:conclusion}

In this study, we present a novel training scheme for average modeling approach voice conversion, that incorporates both frame reconstruction loss and speaker identity cycle consistency loss. The conversion model takes reference speaker embedding from target speech as an auxiliary input. With a pre-trained loss network, a cycle consistency loss is used to minimize the distance between speaker embeddings of reference speech and converted speech. Both objective and subjective evaluation results suggest that the speaker similarity is improved with speaker embedding related cycle consistency loss. 

\section{Acknowledgements}
\label{sec:Acknowledgements}
The work is supported by the National Research Foundation, Singapore under its AI Singapore Programme (AISG Award No: AISG-GC-2019-002 and AISG Award No:  AISG-100E-2018-006). This research is also supported by Human-Robot Interaction Phase 1 (Grant No.: 192 25 00054), National Research Foundation, Singapore under the National Robotics Programme, and Programmatic Grant No.: A18A2b0046 (Human Robot Collaborative AI for AME) from the Singapore Government’s Research, Innovation and Enterprise 2020 plan in the Advanced Manufacturing and Engineering domain.

\bibliographystyle{IEEEbib}
\bibliography{slt}

\end{document}